\documentclass[]{spie}  

\usepackage{etoolbox} 
\usepackage[table]{xcolor}
\usepackage{amssymb}
\usepackage{pifont}
\usepackage{eso-pic}

\usepackage{amsmath,amsfonts,amssymb}
\usepackage{subcaption}
\usepackage{graphicx}
\usepackage{cite} 
\usepackage{times}
\usepackage{epsfig}
\usepackage{amsmath}
\usepackage{nccmath}
\usepackage{amssymb}
\usepackage{mwe}
\usepackage{acro}
\usepackage{amssymb}
\usepackage{xcolor,colortbl}
\usepackage{tabularx}
\usepackage{relsize}
\usepackage{pifont}
\usepackage{booktabs} 
\usepackage{multirow}
\usepackage{multicol}
\usepackage{adjustbox}
\usepackage{float}
\usepackage{placeins}
\usepackage{graphicx}
\usepackage{makecell}
\usepackage{tabu}
\usepackage[colorlinks=true, allcolors=blue]{hyperref}
\usepackage[capitalize]{cleveref}
\usepackage[
  paperwidth=8.5in,
  paperheight=11in,
  textwidth=6.75in,
  textheight=8.75in,
  centering
]{geometry}
\usepackage{eso-pic}
\usepackage{microtype}
\title{StainBridge: Stain-Aware Pairwise Registration of Serial Renal Biopsy Whole-Slide Images Across Structural and Immunohistochemical Stains}

\author[a]{Ellen Wei}
\author[b]{Bohang Jiang}
\author[c]{Yanfan Zhu}
\author[d]{Daniel Reisenbüchler}
\author[e]{Kenji Ikemura}
\author[e]{Steven Salvatore}
\author[e]{Surya Seshan}
\author[e]{Thangamani Muthukumar}
\author[e,f]{Mert R. Sabuncu}
\author[e,g]{Yihe Yang}
\author[e]{Ruining Deng}

\affil[a]{University of California, Los Angeles, Los Angeles, CA, 90095, USA}
\affil[b]{Massachusetts General Brigham, Boston, MA, 02115, USA}
\affil[c]{Vanderbilt University, Nashville, TN, 37235, USA}
\affil[d]{University of Regensburg, Regensburg, Bavaria 93053, DE}
\affil[e]{Weill Medical College of Cornell University, New York, NY 10065, USA}
\affil[f]{Cornell Tech, New York, NY 10044, USA}
\affil[g]{Northwell Health, New Hyde Park, NY 11040, USA}

\authorinfo{Corresponding author: Ruining Deng\\
E-mail: rud4004@med.cornell.edu}

\begin{document} 

\maketitle

\begin{abstract} 
Accurate pairwise registration of serial two-dimensional whole-slide images is a prerequisite for three-dimensional reconstruction of histopathology sections. Existing cross-stain benchmarks have advanced registration of differently stained histology, including structural-to-immunohistochemistry (IHC) registration, but serial renal biopsy stacks remain challenging because they combine multiple structural stains with diverse IHC markers whose expression may be sparse or absent. As a step toward future three-dimensional reconstruction, we present \textit{StainBridge}, a stain-aware framework for pairwise registration of serial renal biopsy whole-slide images across structural and IHC stains. The framework evaluates stain deconvolution, intensity normalization, and tissue-mask injection as complementary preprocessing components, followed by XFeat-based affine alignment and four nonrigid registration approaches. We studied 23 renal biopsy cases comprising 338 whole-slide images, four structural stains, and ten IHC markers. Corresponding functional tissue units were manually annotated across consecutive sections, yielding 1,468 landmark correspondences across 272 landmark-annotated image pairs. Performance was assessed using tissue-mask Dice overlap, functional-unit centroid distance in $\mu\mathrm{m}$, and tissue-restricted structural similarity. 
Nonrigid refinement improved on XFeat's affine-only alignment for three of the four evaluated backends, with only VoxelMorph failing to do so. DeeperHistReg, which performs its own independent initial alignment and does not depend on XFeat, achieved the best overall pooled landmark accuracy and successfully registered a broader set of pairs than the three XFeat-initialized backends, including all attempted structural-IHC pairs. Among the three backends that refine the XFeat affine, preprocessing consistently improved landmark accuracy for ConvexAdam and FireANTs across every stain-pairing category, with FireANTs achieving both the largest single preprocessing benefit in structural-IHC pairs, and the best overall pooled tissue overlap. These findings provide practical guidance for cross-stain registration and establish a foundation for future patient-level three-dimensional reconstruction and integrated analysis of renal tissue architecture and molecular expression.
\end{abstract}


\section{Introduction}  
\label{sec:intro} 

\begin{figure}[htbp]
  \centering
  \includegraphics[width=0.9\textwidth]{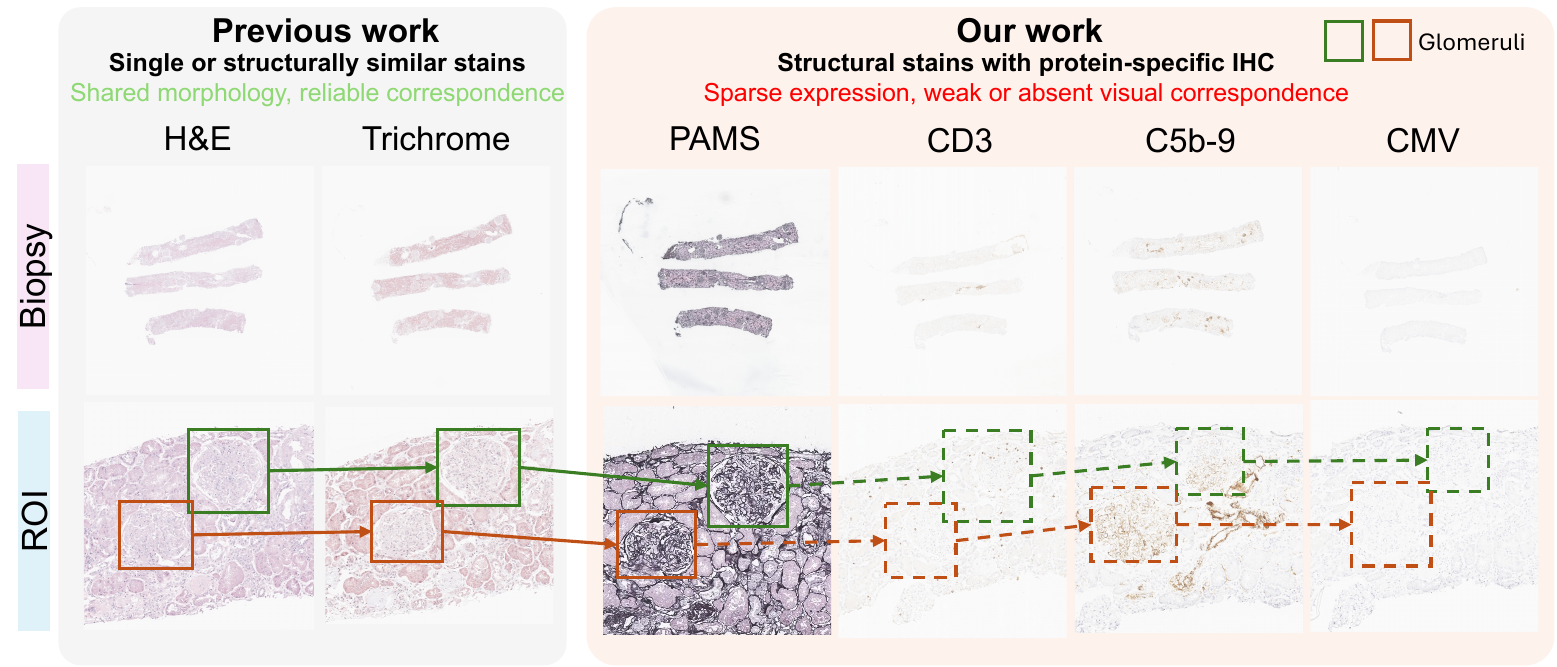}
  \caption{Conceptual comparison of registration settings for serial renal biopsy WSIs. Structurally similar stains, illustrated by H\&E and Trichrome, preserve shared morphology and provide readily visible tissue correspondence. The setting studied here links structural stains such as PAMS with protein-specific IHC stains, including CD3, C5b-9, and CMV, where sparse or absent marker expression can obscure corresponding anatomy. Green and orange boxes track two glomeruli across the serial sections. Solid arrows indicate readily visible correspondence, whereas dashed arrows indicate correspondence across dissimilar stains.}
  \label{fig:initial}
\end{figure}

Three-dimensional (3D) reconstruction of serial histopathology sections can reveal spatial relationships among tissue structures that are inaccessible in any individual two-dimensional whole-slide image (WSI)~\cite{deng2021map3d,deng2022dense,zhao2026mori}. In renal pathology, glomeruli and other functional tissue units extend through tissue depth and may be only partially represented in a single section~\cite{deng2023omni,ancajas2023cellular,ancajas2023harnessing}. Pairwise alignment of serial WSIs is therefore an enabling step for future 3D association of these units and for connecting tissue morphology with marker expression within an individual biopsy~\cite{li2023end,xiong2025zeroreg3d,zhu2025asign,deng2024hats,deng2024prpseg}. The present study evaluates this pairwise registration step rather than a composed 3D reconstruction. Renal biopsy stacks may contain structural stains, such as hematoxylin and eosin (H\&E) and periodic acid--Schiff (PAS), together with protein-specific immunohistochemistry (IHC) stains. The resulting cross-stain correspondence problem is illustrated in Fig.~\ref{fig:initial}.

Serial-section registration must recover spatial correspondence disrupted by tissue
sectioning, mounting, and scanning~\cite{borovec2020anhir,weitz2024acrobat} while
accommodating global displacement, local deformation, missing
tissue~\cite{Wodzinski_2024,Wodzinski2024DeeperHistRegRW} and
stain-dependent appearance~\cite{wodzinski2021deephistreg}. Cross-stain histology registration is an established research problem rather than an unexplored setting. The Automatic Non-Rigid Histological Image Registration (ANHIR) challenge evaluated serial histological sections stained with different dyes~\cite{borovec2020anhir}, and the Automatic Registration of Breast Cancer Tissue (ACROBAT) challenge specifically evaluated H\&E-to-IHC registration in breast cancer specimens~\cite{weitz2024acrobat}. However, these benchmarks do not directly represent serial renal biopsy stacks spanning multiple structural stains and a diverse panel of IHC markers. Renal biopsy sections may contain multiple small tissue cores~\cite{li2023end}, while protein-specific expression may be sparse or absent within corresponding glomeruli~\cite{deng2021map3d,deng2022dense}. The resulting weak or missing appearance cues create a distinct cross-stain registration problem whose response to preprocessing and backend selection has not been systematically characterized in this renal setting.

We present \textit{StainBridge}, a stain-aware framework for pairwise cross-stain registration of serial renal biopsy WSIs. The framework evaluates stain deconvolution~\cite{ruifrok2001quantification,landini2021colourdeconvolution}, intensity normalization, and tissue-mask injection as complementary preprocessing components rather than assuming that each option is required in every configuration. XFeat-based affine registration establishes global correspondence, followed by evaluation of four nonrigid registration backends. Rather than assuming a universal preprocessing configuration or a single optimal backend, StainBridge evaluates how preprocessing benefit varies across stain-pairing categories and registration methods. Performance is assessed using manually matched functional tissue units, tissue-mask overlap, registration error, and pixel-level structural similarity. The primary contribution is a systematic characterization of the interaction among stain type, preprocessing strategy, and registration backend in serial renal biopsy images. This pairwise registration study provides practical guidance and an enabling foundation for future patient-level 3D reconstruction and integrated analysis of renal tissue architecture and molecular expression.

\section{Method}

\begin{figure}[htbp]
  \centering
  \includegraphics[width=0.9\textwidth]{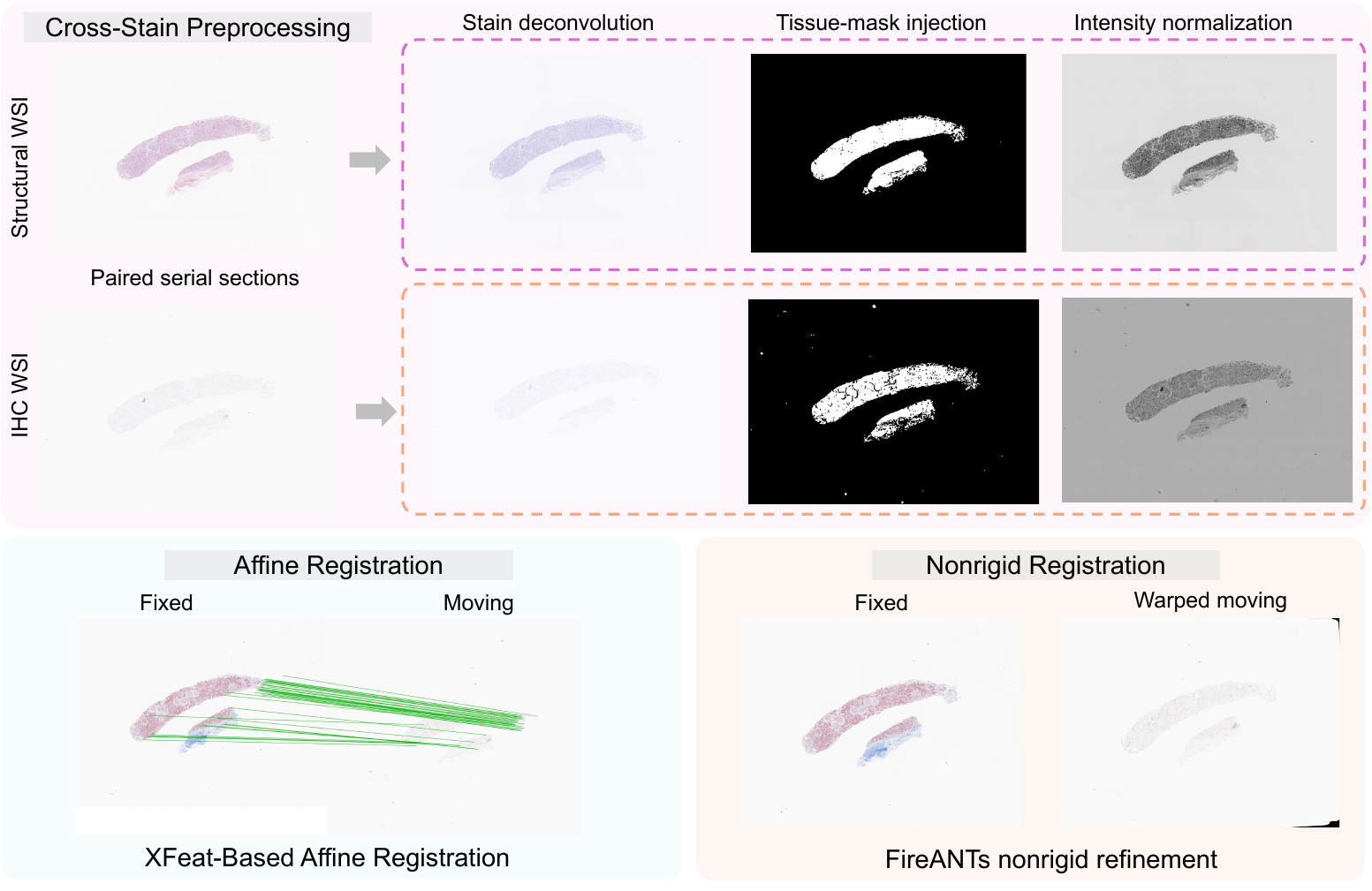}
  \caption{Overview of the cross-stain preprocessing and registration framework. Paired structural and IHC WSIs are processed using stain deconvolution, tissue-mask injection, and intensity normalization, evaluated individually and in combination. XFeat estimates affine alignment. ConvexAdam, FireANTs, and VoxelMorph subsequently estimate nonrigid deformation from the affine-aligned image pairs, whereas DeeperHistReg performs its own initial alignment and nonrigid registration.}
  \label{fig:workflow}
\end{figure}

The framework combines cross-stain preprocessing with affine and nonrigid registration, as illustrated in Fig.~\ref{fig:workflow}. It is designed for paired serial sections in which corresponding tissue structures are preserved but may have substantially different color, intensity, and visible marker expression. The preprocessing components reduce these stain-dependent appearance differences, while the registration pipelines first establish global correspondence and, where applicable, subsequently estimate local tissue deformation.

\subsection{Cross-Stain Preprocessing}
\label{sec:preprocessing}

Cross-stain preprocessing contains three independently configurable components. First, color deconvolution extracts a structurally informative channel using the Ruifrok and Johnston method~\cite{ruifrok2001quantification} and the vector conventions of the ImageJ Colour Deconvolution 2 plugin~\cite{landini2021colourdeconvolution}. This representation emphasizes shared tissue structure while reducing the contribution of stain-specific chromogens; stains without an applicable deconvolution vector use a grayscale fallback. Second, tissue masks identify foreground tissue and support both mask-restricted intensity normalization and optional direct mask injection, in which background pixels are set to zero. These operations reduce the influence of blank slide regions, tissue boundaries, and staining artifacts during correspondence estimation. Third, intensity normalization reduces pairwise intensity variation using robust statistics calculated within the tissue mask. The stain-specific channels, mask-generation procedure, and normalization parameters are provided in Section~\ref{sec:experiment_setting}.

The three components were independently enabled or disabled, producing eight configurations including the raw input. This factorial design allows the individual and combined effects of deconvolution, tissue-mask injection, and intensity normalization to be evaluated. It also permits the preferred preprocessing configuration to vary across stain-pairing categories rather than assuming that a single configuration is optimal for every stain pair or registration method.

\subsection{Registration Pipelines}
\label{sec:registration}

The registration evaluation includes XFeat-based affine registration and four nonrigid registration backends. XFeat first estimates the affine transformation, using the preprocessing configuration selected for each stain-pairing category (Section~\ref{sec:step1}), to recover global translation, rotation, scale, and shear before local deformation is considered. The moving image is then warped using this affine transformation. ConvexAdam, FireANTs, and VoxelMorph use this warped image as their starting point, receiving it under either the raw condition or the stain-deconvolved condition (Section~\ref{sec:step2}), allowing their outputs to be compared from a shared global alignment. DeeperHistReg is evaluated as an independent end-to-end framework because it performs its own initial alignment and cannot directly receive the XFeat transformation; for its preprocessed condition, an externally stain-deconvolved image is supplied first, after which DeeperHistReg applies its own internal grayscale conversion and contrast normalization.

\subsubsection{XFeat-Based Affine Registration}
\label{sec:step1}

For each consecutive pair, the later section was treated as the moving image and the earlier section as the fixed image, providing a consistent direction for composition across the serial stack. XFeat~\cite{potje2024xfeat} was used to extract cross-image correspondences from downsampled registration inputs. Two RANSAC-based homography fits were used only to remove inconsistent matches; the retained correspondences were then used to estimate an affine transformation at the original image resolution. This separation preserves the robustness of projective match filtering while restricting the applied global transformation to an affine model. Pairs that did not satisfy the correspondence criteria were flagged as affine failures. The matching, filtering, and failure parameters are specified in Section~\ref{sec:experiment_setting}.

XFeat was evaluated under all eight preprocessing configurations for structural--structural, structural--IHC, and IHC--IHC pairs. Evaluating the categories separately accounts for differences in the amount and type of shared visual information available for feature matching. Within each category, the configuration with the lowest landmark-weighted centroid distance was selected as the input affine transform for subsequent nonrigid registration, as reported in Table~\ref{tab:xfeat-combo-by-category}. This consistent geometric criterion selected deconvolution with tissue-mask injection for structural--structural pairs, intensity normalization alone for structural--IHC pairs, and deconvolution alone for IHC--IHC pairs.

\subsubsection{Nonrigid Registration Backends}
\label{sec:step2}

Four nonrigid registration backends representing different methodological families were evaluated. For each method, the estimated transformation was mapped to the original image coordinates and applied consistently to the moving image, tissue mask, and glomerular landmarks. The affine-aligned images are subsequently provided to ConvexAdam, FireANTs, and VoxelMorph using either the raw image inputs or stain-deconvolved inputs. Backend-specific settings are reported in Section~\ref{sec:experiment_setting}, and the common evaluation protocol is described in Section~\ref{sec:eval_metrics}.

\noindent\textbf{ConvexAdam.} ConvexAdam~\cite{siebert2025convexadam} performs pair-specific optimization on the XFeat-aligned image pair using a modality independent neighbourhood descriptor with self-similarity context (MIND-SSC)~\cite{heinrich2013towards}. This descriptor emphasizes local structural correspondence and therefore reduces reliance on direct intensity similarity between stains. Because the implementation requires three-dimensional inputs, each 2D image was replicated along a synthetic depth axis, and the central slice of the estimated displacement field was extracted and rescaled to the original image resolution as the 2D deformation.

\noindent\textbf{FireANTs.} FireANTs~\cite{jena2026fireants} performs pair-specific multiscale optimization on the XFeat-aligned image pair using local normalized cross-correlation. A coarse-to-fine schedule progressively estimates global and local deformation, while warp and gradient smoothing regularize the displacement field. The optimized field was then rescaled to the original image resolution before application.

\noindent\textbf{VoxelMorph.} VoxelMorph~\cite{balakrishnan2019voxelmorph} uses a pretrained three-dimensional deep registration network and was included as an off-the-shelf model trained outside the histology domain. The selected checkpoint was trained on brain magnetic resonance images and was evaluated without task-specific fine-tuning. The XFeat-aligned 2D images were resized and replicated along a synthetic depth axis, after which the central slice of the predicted displacement field was extracted and rescaled as the 2D deformation.

\noindent\textbf{DeeperHistReg.} DeeperHistReg~\cite{Wodzinski2024DeeperHistRegRW} combines its own feature-based initial alignment with instance-optimized nonrigid refinement. Its initialization evaluates both conventional and learned feature-matching candidates, followed by multiscale refinement using local normalized cross-correlation and deformation regularization. Because it does not accept the XFeat affine transformation as an external input, it was evaluated as an independent end-to-end pipeline. It is not conditioned on XFeat success and preserves its ability to process pairs for which XFeat initialization fails. For its preprocessed condition, an externally stain-deconvolved image is supplied as input, after which DeeperHistReg applies its own internal grayscale conversion and contrast normalization.

\section{Data \& Experiments}  
\subsection{Data}

The dataset comprises 338 WSIs (7--17 per case) from 23 renal biopsies processed at Weill Medical College of Cornell University. Stains include four structural stains: H\&E, PAS, PAMS, Masson's trichrome, and ten IHC markers: BKV, C4d, C5b-9, CD3, CD20, CD56, CMV, FoxP3, Granzyme, and KP1. Slides were digitized on an Aperio GT450 (v1.3.3) at $40\times$ magnification. QuPath matching yielded 1,468 glomerular landmark correspondences across 272 consecutive pairs; 43 pairs without persistent visible glomeruli were excluded only from landmark IoU and centroid-distance evaluation. The 315 pairs comprise 181 structural--structural, 22 structural--IHC, and 112 IHC--IHC pairs; Direction was not analyzed. DeeperHistReg used 300 pairs after one case was excluded (Section~\ref{sec:results_deeperhistreg}).

\subsection{Experiment Setting}
\label{sec:experiment_setting}

\noindent\textbf{Preprocessing.} Color deconvolution retained the hematoxylin channel for H\&E and vector-compatible DAB IHC stains (CD3, KP1, CD56, Granzyme, CD20, and BKV), the hematoxylin channel from a generic hematoxylin/PAS vector for PAS, and the methyl blue channel from a generic methyl blue/Ponceau-fuchsin vector for Trichrome. Because stain-specific controls were unavailable, generic literature-derived vectors were used. PAMS, C5b-9, C4d, FoxP3, and CMV used ITU-R BT.601 grayscale conversion, $0.299R+0.587G+0.114B$. Tissue masks were generated at $1/8$ resolution using per-image Otsu thresholds on HSV saturation and value, followed by morphological closing (full-resolution-equivalent radius, $3~\text{px}$) and removal of objects and holes smaller than $500~\text{px}^{2}$. Masks were upsampled and rebinarized. For normalization, masks were eroded by $5~\text{px}$; the 0.5th--99.5th percentile range and corresponding trimmed mean and standard deviation were used to stretch and recenter intensities toward mean 0.5 and standard deviation 0.25 in $[0,1]$. For mask injection, pixels outside the non-eroded mask were set to zero.

\noindent\textbf{Affine Registration and Configuration Selection.} XFeat retained the top $k=4096$ correspondences from inputs with maximum dimension $1000~\text{px}$. Of the $N_0$ matches, USAC\_MAGSAC retained $N_1$ using a $3.5~\text{px}$ threshold, 0.999 confidence, and 1000 iterations; RANSAC retained $N_2$ using a $5.0~\text{px}$ threshold. No affine was saved for $N_1<20$; a flagged identity fallback was specified for $N_2\leq10$, although none occurred. Otherwise, full-resolution correspondences were passed to OpenCV \texttt{estimateAffine2D} with a $3~\text{px}$ RANSAC threshold, 2000 iterations, 0.99 confidence, and 10 refinement iterations. All eight preprocessing combinations were evaluated separately for the three stain-pairing categories. Within each category, the affine configuration minimizing landmark-weighted centroid distance was selected, with Dice as a secondary criterion; SSIM was excluded because it measures appearance rather than geometry. Separately, pooled centroid distance selected stain deconvolution as the preprocessed nonrigid input. The category-selected affine was applied first, after which the affine-aligned pair was supplied to ConvexAdam, FireANTs, or VoxelMorph under either the raw or stain-deconvolved condition.

\noindent\textbf{Nonrigid Registration.} ConvexAdam (v0.2.0) used \path{convex_adam_pt} with \path{mind_r=1}, \path{mind_d=2}, \path{lambda_weight=1.25}, \path{grid_sp=4}, \path{disp_hw=8}, \path{selected_niter=120}, \path{selected_smooth=3}, \path{grid_sp_adam=2}, and inverse consistency. Inputs were limited to $800~\text{px}$ and replicated across 24 slices; the central field was converted to $(dx,dy)$ and rescaled. FireANTs (v1.5.1) used \texttt{GreedyRegistration}, local normalized cross-correlation with kernel 7, scales $[4,2,1]$, iterations $[100,50,25]$, warp smoothing $\sigma=0.5$, and gradient smoothing $\sigma=1.0$; inputs were limited to $800~\text{px}$ and the field was rescaled. VoxelMorph (v0.2) used \path{vxm_dense_brain_T1_3D_mse.h5}; images were resized to $160\times192$ and replicated across 224 slices, and the central field was converted to $(dx,dy)$ and rescaled before applying the $200~\text{px}$ full-resolution displacement cap. DeeperHistReg (v1.0.1) used \texttt{default\_initial\_nonrigid} at 20\% resolution. SIFT--RANSAC and SuperPoint~\cite{detone2018superpoint}--SuperGlue~\cite{sarlin2019superglue} candidates were evaluated over four rotations and sizes $150$--$500~\text{px}$ in $50~\text{px}$ increments; the rigid candidate with most inliers was selected, with identity as fallback. Its nonrigid refinement used an eight-level pyramid, local normalized cross-correlation, diffusive regularization, 100 iterations per level, and 200 at the finest level; the saved $2\times H\times W$ field was transposed to $H\times W\times2$. All fields were resized using bilinear interpolation and applied with bilinear interpolation for image content and nearest-neighbor interpolation for tissue masks and landmark annotations.

\begin{table}[htbp]
\centering
\caption{Approximate per-pair registration runtime, excluding preprocessing.}
\label{tab:runtime}
\adjustbox{max width=\textwidth}{
\begin{tabular}{llc}
\toprule
Method & Input & Runtime (s) \\
\midrule
ConvexAdam~\cite{siebert2025convexadam} & Raw / Deconvolved & 47.1 / 48.4 \\
FireANTs~\cite{jena2026fireants} & Raw / Deconvolved & 45.2 / 44.8 \\
VoxelMorph~\cite{balakrishnan2019voxelmorph} & Raw / Deconvolved & 51.4 / 51.5 \\
DeeperHistReg~\cite{Wodzinski2024DeeperHistRegRW} & Raw / Deconvolved & 140.3 / 93.6 \\
\bottomrule
\end{tabular}
}
\end{table}

Table~\ref{tab:runtime} reports approximate wall-clock time measured with Python \texttt{time.time()} around each registration call. GPU operations were not explicitly synchronized, but transferring each field to NumPy before stopping the timer waited for pending computation.

\subsection{Inference and Post-Processing}

ConvexAdam, FireANTs, and VoxelMorph refined the category-selected XFeat affine, whereas DeeperHistReg used its own initial alignment. Each displacement field transformed the moving image, tissue mask, and landmarks into the fixed-image frame. If initialization or a required input was unavailable, the affected slide started a new registration chain and the event was recorded. Registered color images were retained for qualitative review.

\subsection{Evaluation Metrics}
\label{sec:eval_metrics}

Glomeruli were annotated as polygons in QuPath, rasterized, represented by mask centroids, matched by name, and evaluated when present on both slides. Distances used the fixed-image resolution and are reported in micrometres ($\mu\text{m}$). Tissue overlap was measured as $\text{Dice}=2|A\cap B|/(|A|+|B|)$ between fixed and warped moving tissue masks, and landmark overlap as $\text{IoU}=|A\cap B|/|A\cup B|$ between corresponding glomerular masks; both were defined as 1 when both masks were empty and 0 when only one was empty. XFeat inlier ratio was $N_2/N_0$, and landmark coverage was the fraction of registered pairs containing at least one matched glomerulus. Centroid distance and landmark IoU use landmark-weighted means, whereas pair-level Dice and SSIM use pair-weighted means. SSIM~\cite{wang2004image} used scikit-image's $7\times7$ uniform window on grayscale images (range 0--255), averaging its map within the fixed tissue mask. Raw and preprocessed conditions used raw-grayscale and stain-deconvolved representations, respectively; DeeperHistReg outputs were converted after internal preprocessing. SSIM was secondary and was not compared directly across backends.

\section{Results}

The results address two questions: whether cross-stain preprocessing improves registration across methods and how registration behavior changes among structural--structural, structural--IHC, and IHC--IHC pairs. Table~\ref{tab:overview} summarizes pooled affine and nonrigid performance, while Tables~\ref{tab:xfeat-combo-by-category} and~\ref{tab:nonrigid_by_category} provide the stain-pairing breakdowns.

\subsection{Registration Results with Cross-Stain Preprocessing}
\label{sec:results_preprocessing}

\subsubsection{Affine Registration}

Across all stain pairings, XFeat preprocessing reduced landmark-weighted centroid distance from 145.1~$\mu\text{m}$ on raw images to 118.9~$\mu\text{m}$ with deconvolution, while Dice increased from 0.679 to 0.686. The highest pooled Dice, 0.699, was obtained by combining deconvolution with tissue-mask injection, but this configuration did not minimize centroid distance. Other combinations improved overlap without improving landmark accuracy, and the most complex configuration was not consistently superior. Cross-stain preprocessing therefore benefited affine registration, but the optimal components depended on the metric and stain-pairing category rather than following a single universal combination.

\begin{table}
\centering
\caption{Registration performance by method and input condition. Deconv.: stain deconvolution; Mask: tissue-mask injection; Norm.: intensity normalization; CD: centroid distance; SSIM: structural similarity index measure. For nonrigid methods, Deconv. denotes an additionally deconvolved affine-aligned input. Bold indicates the best value within each registration type.}
\label{tab:overview}
\adjustbox{max width=0.8\textwidth}{%
\begin{tabular}{llcccccc}
\toprule
Type & Method & Deconv. & Mask & Norm. & Dice & CD* ($\mu\text{m}$) & SSIM \\
\midrule
\multirow{8}{*}{Affine}
& XFeat~\cite{potje2024xfeat} &  &  &  & 0.679 & 145.1 & 0.237 \\
& XFeat~\cite{potje2024xfeat} & \checkmark &  &  & 0.686 & \textbf{118.9} & \textbf{0.278} \\
& XFeat~\cite{potje2024xfeat} & \checkmark & \checkmark &  & \textbf{0.699} & 123.7 & 0.132 \\
& XFeat~\cite{potje2024xfeat} & \checkmark &  & \checkmark & 0.688 & 134.1 & 0.042 \\
& XFeat~\cite{potje2024xfeat} & \checkmark & \checkmark & \checkmark & 0.697 & 134.5 & 0.019 \\
& XFeat~\cite{potje2024xfeat} &  & \checkmark &  & 0.695 & 128.5 & 0.116 \\
& XFeat~\cite{potje2024xfeat} &  &  & \checkmark & 0.690 & 144.2 & 0.044 \\
& XFeat~\cite{potje2024xfeat} &  & \checkmark & \checkmark & 0.688 & 163.5 & 0.021 \\
\midrule
\multirow{8}{*}{Nonrigid}
& FireANTs~\cite{jena2026fireants} &  &  &  & 0.770 & 125.5 & 0.247 \\
& FireANTs~\cite{jena2026fireants} & \checkmark & &
& \textbf{0.779} & 81.3 & 0.286 \\
& VoxelMorph~\cite{balakrishnan2019voxelmorph} &  &  &  & 0.682 & 225.1 & 0.246 \\
& VoxelMorph~\cite{balakrishnan2019voxelmorph} & \checkmark & &
& 0.670 & 201.3 & \textbf{0.294} \\
& ConvexAdam~\cite{siebert2025convexadam} &  &  &  & 0.760 & 111.1 & 0.279 \\
& ConvexAdam~\cite{siebert2025convexadam} & \checkmark & & 
& 0.770 & 72.8 & 0.281 \\
& DeeperHistReg~\cite{Wodzinski2024DeeperHistRegRW} &  &  &  & 0.777 & \textbf{46.9} & 0.251 \\
& DeeperHistReg~\cite{Wodzinski2024DeeperHistRegRW} & \checkmark & &
& 0.775 & 51.3 & 0.271 \\
\bottomrule
\vspace{2pt}
{\footnotesize *Landmark-count-weighted.}
\end{tabular}%
}
\end{table}

\subsubsection{Nonrigid Registration}
\label{sec:backend_results}

The effect of deconvolution was evaluated jointly across the nonrigid methods rather than by treating each backend as a separate experiment. Among the three XFeat-initialized backends, preprocessing reduced pooled centroid distance for every method, with reductions ranging from 10.6\% to 35.2\%. Dice also increased for two of these three methods, although VoxelMorph showed a small decrease from 0.682 to 0.670. DeeperHistReg behaved differently: its already low raw-input centroid distance of 46.9~$\mu\text{m}$ increased to 51.3~$\mu\text{m}$ after external deconvolution, consistent with its use of independent initialization and internal grayscale and contrast normalization. Overall, preprocessing improved geometric accuracy for the XFeat-initialized group but did not provide a universal benefit to every complete pipeline. Nonrigid refinement outperformed the best pooled affine centroid distance for ConvexAdam, FireANTs, and DeeperHistReg, whereas VoxelMorph remained less accurate than affine-only registration.

\subsection{Registration Across Stain-Pairing Categories}
\label{sec:results_deeperhistreg}

The affine results show that stain pairing determined which preprocessing components were useful (Table~\ref{tab:xfeat-combo-by-category}). Structural--structural pairs had the strongest tissue overlap and benefited most from deconvolution with mask injection, which reduced centroid distance from 136.8 to 99.4~$\mu\text{m}$. Structural--IHC pairs responded differently: intensity normalization alone produced the best Dice (0.664) and centroid distance (92.4~$\mu\text{m}$), whereas deconvolution alone increased the error to 156.3~$\mu\text{m}$. IHC--IHC pairs had the weakest affine Dice but improved with deconvolution, which reduced centroid distance from 161.5 to 122.4~$\mu\text{m}$. These results show that preprocessing should be selected by stain-pairing category rather than pooled performance alone.

\begin{figure}
  \centering
  \includegraphics[width=0.9\textwidth]{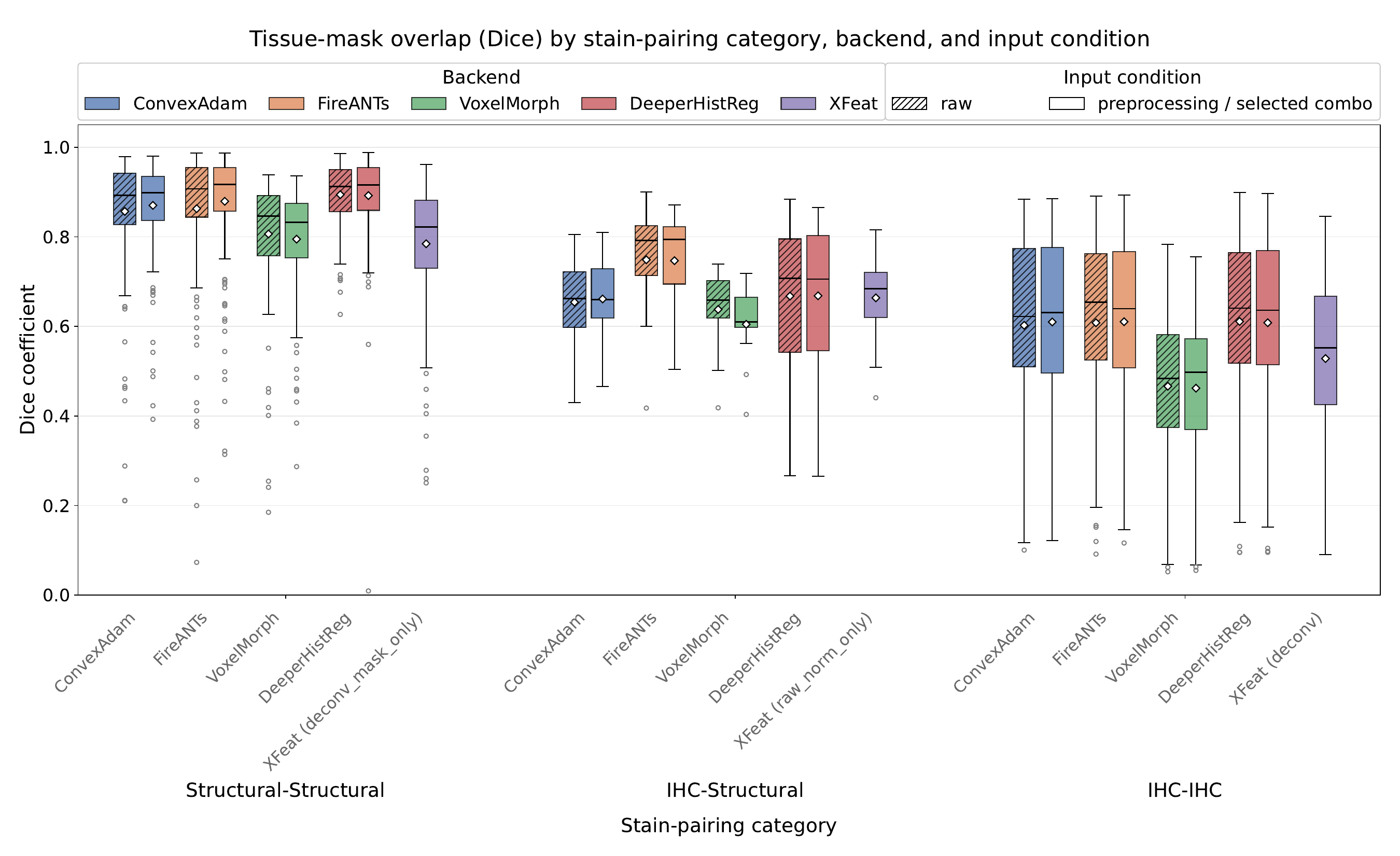}
  \caption{Tissue-mask Dice distributions by stain-pairing category, registration backend, and input condition. Hatched boxes represent raw inputs, solid boxes represent preprocessed inputs or the selected XFeat combination, and diamonds indicate means. Higher values indicate better tissue overlap.}
  \label{fig:dice_by_category}
\end{figure}

\begin{figure}
  \centering
  \includegraphics[width=0.9\textwidth]{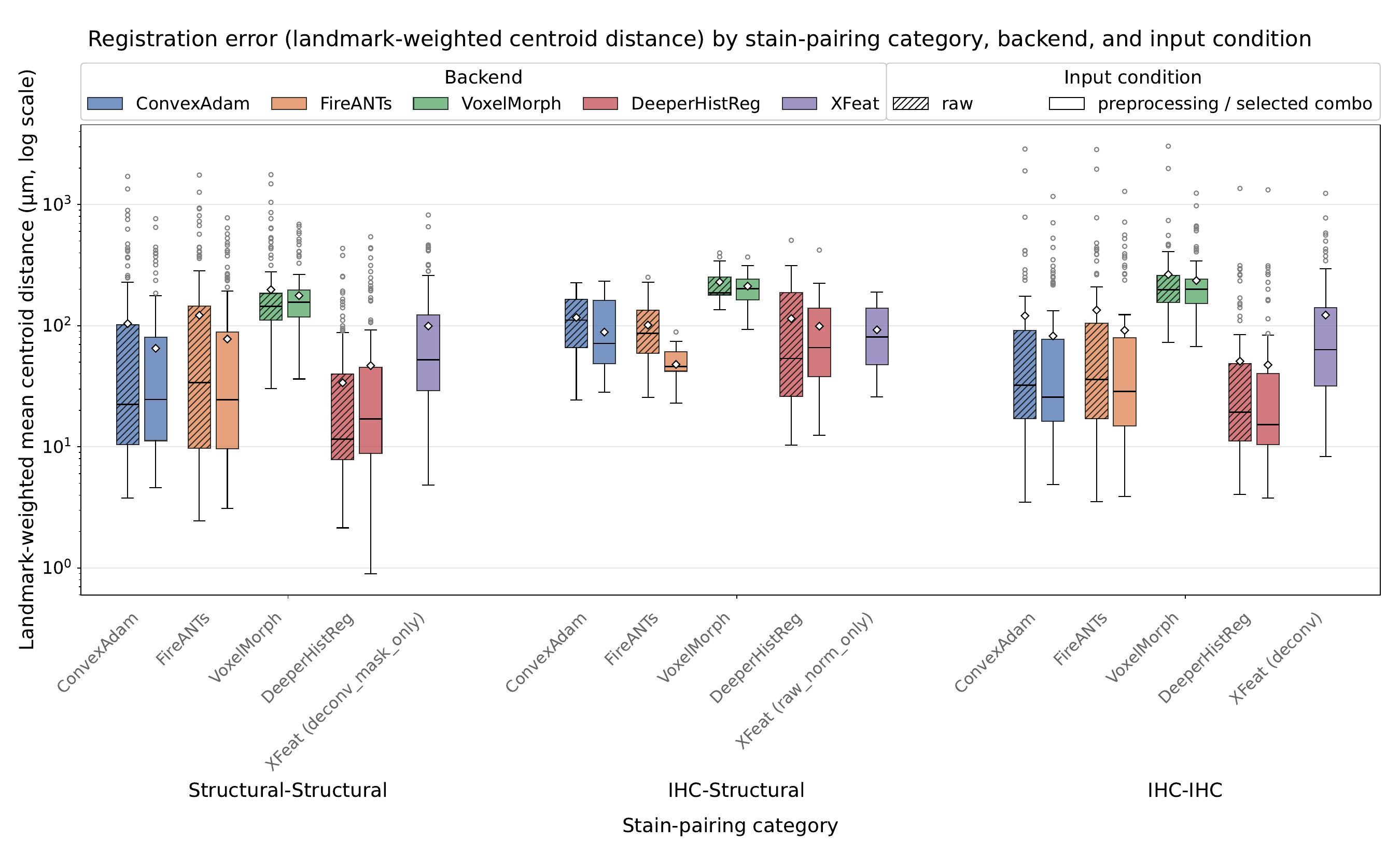}
  \caption{Landmark-weighted centroid-distance distributions by stain-pairing category, registration backend, and input condition. Hatched boxes represent raw inputs, solid boxes represent preprocessed inputs or the selected XFeat combination, and diamonds indicate means. Distances are shown in $\mu\text{m}$ on a logarithmic scale; lower values indicate better alignment.}
  \label{fig:centroid_by_category}
\end{figure}

\begin{figure}
  \centering
  \includegraphics[width=0.9\textwidth]{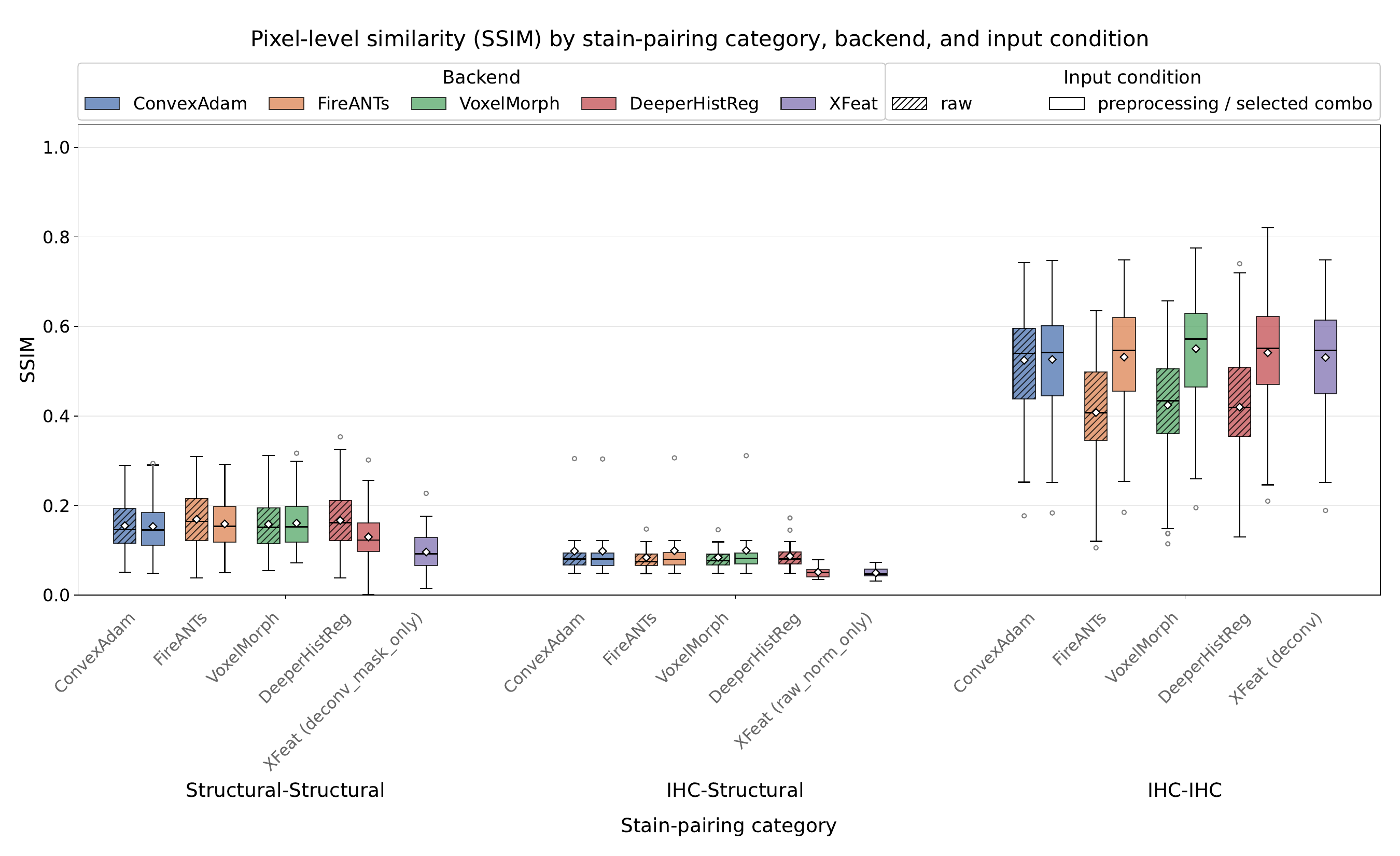}
  \caption{Tissue-restricted SSIM distributions by stain-pairing category, registration backend, and input condition. Hatched boxes represent raw inputs, solid boxes represent preprocessed inputs or the selected XFeat combination, and diamonds indicate means. Higher values indicate greater pixel-level appearance similarity.}
  \label{fig:ssim_by_category}
\end{figure}

The nonrigid results followed related category-level trends (Table~\ref{tab:nonrigid_by_category}). Structural--structural pairs remained easiest by tissue overlap, with Dice reaching 0.894, whereas IHC--IHC pairs had lower Dice despite comparatively high SSIM. Structural--IHC pairs showed the weakest appearance similarity and most fragile initialization: the XFeat-dependent backends registered 13 of 22 pairs, whereas DeeperHistReg registered all 21 available pairs. This category also produced the largest preprocessing benefit, a 52.7\% centroid-distance reduction. Across the three XFeat-initialized backends, preprocessing reduced centroid distance in every category, with improvements of 10.7\%--37.6\% for structural--structural, 7.5\%--52.7\% for structural--IHC, and 10.9\%--32.0\% for IHC--IHC pairs. DeeperHistReg improved in both categories involving IHC but worsened for structural--structural pairs. Dice, centroid-distance, and SSIM trends are illustrated in Figs.~\ref{fig:dice_by_category},~\ref{fig:centroid_by_category}, and~\ref{fig:ssim_by_category}, respectively.

\begin{table}
\centering
\caption{XFeat~\cite{potje2024xfeat} performance across preprocessing combinations and stain-pairing groups. S: structural stain; Deconv.: stain deconvolution; Norm.: intensity normalization; IR: inlier ratio; IoU: intersection over union; CD: centroid distance. Overall (pooled) combines all groups, and bold indicates the best value within each group.}
\label{tab:xfeat-combo-by-category}
\adjustbox{max width=0.5\textwidth}{%
\begin{tabular}{lcccccccc}
\toprule
Stain Pairing & Deconv. & Mask & Norm. & Dice & IR & IoU* & CD* ($\mu\text{m}$) \\
\midrule
\multirow{8}{*}{S--S}
&  &  &  & 0.773 & 0.277 & 0.528 & 136.8 \\
& \checkmark &  &  & 0.783 & 0.261 & 0.536 & 113.1 \\
& \checkmark & \checkmark &  & 0.784 & 0.337 & 0.535 & \textbf{99.4} \\
& \checkmark &  & \checkmark & \textbf{0.793} & 0.276 & 0.548 & 101.9 \\
& \checkmark & \checkmark & \checkmark
& 0.783 & \textbf{0.350} & \textbf{0.564} & 101.1 \\
&  & \checkmark &  & 0.783 & 0.338 & 0.534 & 105.3 \\
&  &  & \checkmark & 0.786 & 0.306 & 0.545 & 125.6 \\
&  & \checkmark & \checkmark & 0.778 & 0.342 & 0.519 & 112.1 \\
\midrule
\multirow{8}{*}{S--IHC}
&  &  &  & 0.663 & 0.081 & 0.345 & 117.8 \\
& \checkmark &  &  & 0.609 & 0.104 & 0.373 & 156.3 \\
& \checkmark & \checkmark &  & 0.636 & 0.163 & 0.424 & 127.1 \\
& \checkmark &  & \checkmark
& 0.634 & 0.105 & \textbf{0.427} & 114.1 \\
& \checkmark & \checkmark & \checkmark
& 0.646 & \textbf{0.174} & 0.390 & 164.3 \\
&  & \checkmark &  & 0.635 & 0.167 & 0.417 & 159.1 \\
&  &  & \checkmark
& \textbf{0.664} & 0.085 & 0.412 & \textbf{92.4} \\
&  & \checkmark & \checkmark & 0.654 & 0.173 & 0.426 & 130.2 \\
\midrule
\multirow{8}{*}{IHC--IHC}
&  &  &  & 0.516 & 0.185 & 0.468 & 161.5 \\
& \checkmark &  &  &
0.528 & \textbf{0.260} & \textbf{0.487} & \textbf{122.4} \\
& \checkmark & \checkmark &  &
\textbf{0.548} & 0.210 & 0.443 & 161.0 \\
& \checkmark &  & \checkmark & 0.516 & 0.232 & 0.479 & 184.8 \\
& \checkmark & \checkmark & \checkmark
& 0.541 & 0.198 & 0.389 & 181.9 \\
&  & \checkmark &  & 0.539 & 0.212 & 0.406 & 160.2 \\
&  &  & \checkmark & 0.526 & 0.185 & 0.469 & 179.1 \\
&  & \checkmark & \checkmark & 0.531 & 0.207 & 0.402 & 246.4 \\
\midrule
\multirow{8}{*}{Overall (pooled)}
&  &  &  & 0.679 & 0.231 & 0.495 & 145.1 \\
& \checkmark &  &  &
0.686 & 0.250 & 0.509 & \textbf{118.9} \\
& \checkmark & \checkmark &  &
\textbf{0.699} & 0.280 & 0.495 & 123.7 \\
& \checkmark &  & \checkmark &
0.688 & 0.249 & \textbf{0.515} & 134.1 \\
& \checkmark & \checkmark & \checkmark &
0.697 & \textbf{0.284} & 0.488 & 134.5 \\
&  & \checkmark &  & 0.695 & 0.281 & 0.480 & 128.5 \\
&  &  & \checkmark & 0.689 & 0.247 & 0.509 & 144.2 \\
&  & \checkmark & \checkmark & 0.688 & 0.282 & 0.469 & 163.5 \\
\bottomrule
\vspace{2pt}
{\footnotesize *Landmark-count-weighted.}
\end{tabular}%
}
\end{table}

\begin{table}
\centering
\caption{Nonrigid registration performance by input condition and stain-pairing category. Preprocessed denotes a deconvolved input after category-specific affine alignment; DeeperHistReg uses independent initialization. CD: centroid distance; SSIM: structural similarity index measure. Bold indicates the best value within each category; an italicized bold value identifies a DeeperHistReg best, with the best XFeat-initialized result also bolded.}
\label{tab:nonrigid_by_category}
\adjustbox{max width=0.5\textwidth}{%
\begin{tabular}{llcccc}
\toprule
Category & Method & Preprocessed & Dice & CD* ($\mu\text{m}$) & SSIM \\
\midrule
\multirow{8}{*}{Pooled}
& \multirow{2}{*}{FireANTs~\cite{jena2026fireants}} &  & 0.770 & 125.5 & 0.247 \\
& & \checkmark & \textbf{0.779} & 81.3 & 0.286 \\
\cmidrule(lr){2-6}
& \multirow{2}{*}{VoxelMorph~\cite{balakrishnan2019voxelmorph}} &  & 0.682 & 225.1 & 0.246 \\
& & \checkmark & 0.670 & 201.3 & \textbf{0.294} \\
\cmidrule(lr){2-6}
& \multirow{2}{*}{ConvexAdam~\cite{siebert2025convexadam}} &  & 0.760 & 111.1 & 0.279 \\
& & \checkmark & 0.770 & \textbf{72.8} & 0.281 \\
\cmidrule(lr){2-6}
& \multirow{2}{*}{DeeperHistReg~\cite{Wodzinski2024DeeperHistRegRW}} &  & 0.777 & \textbf{\textit{46.9}} & 0.251 \\
& & \checkmark & 0.775 & 51.3 & 0.271 \\
\midrule
\multirow{8}{*}{Structural--structural}
& \multirow{2}{*}{FireANTs~\cite{jena2026fireants}} &  & 0.863 & 121.9 & \textbf{0.169} \\
& & \checkmark & \textbf{0.879} & 77.6 & 0.159 \\
\cmidrule(lr){2-6}
& \multirow{2}{*}{VoxelMorph~\cite{balakrishnan2019voxelmorph}} &  & 0.806 & 198.3 & 0.158 \\
& & \checkmark & 0.795 & 177.1 & 0.161 \\
\cmidrule(lr){2-6}
& \multirow{2}{*}{ConvexAdam~\cite{siebert2025convexadam}} &  & 0.857 & 104.1 & 0.155 \\
& & \checkmark & 0.870 & \textbf{65.0} & 0.153 \\
\cmidrule(lr){2-6}
& \multirow{2}{*}{DeeperHistReg~\cite{Wodzinski2024DeeperHistRegRW}} & & \textbf{\textit{0.894}} & \textbf{\textit{33.8}} & 0.167 \\
& & \checkmark & 0.892 & 46.8 & 0.130 \\
\midrule
\multirow{8}{*}{Structural--IHC}
& \multirow{2}{*}{FireANTs~\cite{jena2026fireants}} &  & \textbf{0.749} & 101.1 & 0.084 \\
& & \checkmark & 0.747 & \textbf{47.9} & 0.099 \\
\cmidrule(lr){2-6}
& \multirow{2}{*}{VoxelMorph~\cite{balakrishnan2019voxelmorph}} &  & 0.638 & 229.2 & 0.084 \\
& & \checkmark & 0.605 & 212.0 & \textbf{0.100} \\
\cmidrule(lr){2-6}
& \multirow{2}{*}{ConvexAdam~\cite{siebert2025convexadam}} &  & 0.654 & 116.9 & 0.098 \\
& & \checkmark & 0.661 & 88.5 & 0.098 \\
\cmidrule(lr){2-6}
& \multirow{2}{*}{DeeperHistReg~\cite{Wodzinski2024DeeperHistRegRW}} &  & 0.667 & 114.5 & 0.087 \\
& & \checkmark & 0.668 & 99.1 & 0.051 \\
\midrule
\multirow{8}{*}{IHC--IHC}
& \multirow{2}{*}{FireANTs~\cite{jena2026fireants}} &  & 0.608 & 134.5 & 0.407 \\
& & \checkmark & \textbf{0.610} & 91.5 & 0.531 \\
\cmidrule(lr){2-6}
& \multirow{2}{*}{VoxelMorph~\cite{balakrishnan2019voxelmorph}} &  & 0.466 & 264.8 & 0.424 \\
& & \checkmark & 0.462 & 236.0 & \textbf{0.550} \\
\cmidrule(lr){2-6}
& \multirow{2}{*}{ConvexAdam~\cite{siebert2025convexadam}} &  & 0.602 & 120.8 & 0.524 \\
& & \checkmark & 0.610 & \textbf{82.2} & 0.526 \\
\cmidrule(lr){2-6}
& \multirow{2}{*}{DeeperHistReg~\cite{Wodzinski2024DeeperHistRegRW}} &  & \textbf{\textit{0.611}} & 50.9 & 0.420 \\
& & \checkmark & 0.608 & \textbf{\textit{47.5}} & 0.541 \\
\bottomrule
\multicolumn{6}{l}{\footnotesize *Landmark-count-weighted.} \\
\end{tabular}%
}
\end{table}

\FloatBarrier

\subsection{Discussion, Limitations, and Future Work}
\label{sec:limitations}
Cross-stain preprocessing was most reliable for the three backends that refined the same XFeat affine initialization, with the strongest geometric benefit in structural--IHC pairs; however, the optimal components varied by stain pairing and metric. External preprocessing did not consistently improve DeeperHistReg, likely because it uses independent initialization and internal grayscale and contrast normalization. The backend results therefore represent practical tradeoffs rather than a universal winner: FireANTs and ConvexAdam showed complementary category-level strengths, while DeeperHistReg provided broader coverage and stronger landmark accuracy in structural--structural and IHC--IHC pairs. Because DeeperHistReg used a different initialization and was evaluated on 22 cases after one case was excluded for repeated out-of-memory failures, its comparison with the XFeat-initialized backends was not strictly matched; both coverage and geometric accuracy should be considered when selecting a pipeline.

This study is limited by its 23-case, single-center cohort and the relatively small structural--IHC subset. Deconvolution vectors were unavailable for C5b-9, FoxP3, C4d, and CMV, which used a grayscale fallback, and whole-slide registration did not include an independent check of slide content or separately model multiple tissue cores. In addition, the study evaluates pairwise two-dimensional registration rather than a composed three-dimensional reconstruction. Future work should validate stain-specific deconvolution vectors, compare all backends on exactly matched pairs, improve initialization and failure detection, test multicenter cohorts, and compose the pairwise transformations into patient-level three-dimensional reconstructions.

\section{New or Breakthrough Work to be Presented}
This work presents a systematic evaluation of pairwise registration for serial renal biopsy whole-slide images spanning four structural stains and ten immunohistochemistry (IHC) markers. Unlike ANHIR and ACROBAT, which established broader differently stained and H\&E-to-IHC benchmarks, respectively, this study focuses on renal biopsies with multiple small tissue cores, sparse or absent marker expression, and glomerulus-based evaluation. Its principal contribution is not a new registration algorithm, but a controlled characterization of how stain-aware preprocessing interacts with affine and nonrigid registration across structural--structural, structural--IHC, and IHC--IHC pairs. The results show that preprocessing generally improves XFeat-initialized registration, although its magnitude and optimal components vary by stain pairing, while independent initialization can substantially increase coverage when cross-stain appearance cues are weak.

\section{Conclusion}
We presented StainBridge, a stain-aware framework for pairwise registration of serial renal biopsy whole-slide images across structural and immunohistochemistry stains. Cross-stain preprocessing improved pooled affine registration and reduced centroid distance for all three XFeat-initialized nonrigid backends, although its effect varied by stain pairing and did not consistently benefit DeeperHistReg, which uses independent initialization and internal preprocessing. Structural--structural pairs achieved the strongest tissue overlap, structural--IHC pairs showed the greatest initialization fragility and largest preprocessing benefit, and IHC--IHC pairs retained lower tissue overlap despite comparatively high appearance similarity. DeeperHistReg provided the broadest coverage and strongest overall landmark accuracy, whereas the XFeat-initialized methods showed method-dependent tradeoffs. These findings provide practical guidance for cross-stain renal registration and support future validation and patient-level three-dimensional reconstruction.

\section{ACKNOWLEDGMENTS} 
This research was supported by the WCM Radiology AIMI Fellowship, WCM CTSC 2027 Pilot Award, NIH 1R01HL174863-01A1(Sabuncu), and 1U54DK144866-01(Sabuncu).

\bibliography{report} 
\bibliographystyle{spiebib} 

\end{document}